\documentclass[twocolumn,showpacs,amsmath,amssymb,superscriptaddress,prl,aps,final,10pt]{revtex4-1}

\usepackage{graphicx}
\usepackage{dcolumn} 
\usepackage{bm}      
\usepackage{nicefrac}
\usepackage{mathrsfs}
\usepackage[dvips]{color} 
\definecolor{orange}{rgb}{1.0,0.549,0} 

\newcommand{\BACSO}{BaCu$_2$Si$_2$O$_7$}
\newcommand{\BASGE}{Ba\-Cu$_2$\-(Si$_{0.5}$\-Ge$_{0.5})_2$\-O$_7$}
\newcommand{\diff}{\mathrm{d}}

\begin{document}

\title{Distribution of NMR relaxations in a random Heisenberg chain}
\author{T.\ Shiroka}
\author{F.\ Casola}
\email{fcasola@phys.ethz.ch}
\affiliation{Laboratorium f\"ur Festk\"orperphysik, ETH H\"onggerberg, CH-8093 Z\"urich, Switzerland}
\affiliation{Paul Scherrer Institut, CH-5232 Villigen PSI, Switzerland}
\author{V.\ Glazkov}
\affiliation{Laboratorium f\"ur Festk\"orperphysik, ETH
H\"onggerberg, CH-8093 Z\"urich, Switzerland} \affiliation{P.-L.\ 
Kapitza Institute for Physical Problems RAS, 119334 Moscow, Russia}
\author{A.\ Zheludev}
\affiliation{Laboratorium f\"ur Festk\"orperphysik, ETH H\"onggerberg, CH-8093 Z\"urich, Switzerland}
\author{K.\ Pr\v{s}a}
\affiliation{Laboratorium f\"ur Festk\"orperphysik, ETH H\"onggerberg, CH-8093 Z\"urich, Switzerland}
\affiliation{Paul Scherrer Institut, CH-5232 Villigen PSI, Switzerland}
\author{H.-R.\ Ott}
\affiliation{Laboratorium f\"ur Festk\"orperphysik, ETH H\"onggerberg, CH-8093 Z\"urich, Switzerland}
\author{J.\ Mesot}
\affiliation{Laboratorium f\"ur Festk\"orperphysik, ETH H\"onggerberg, CH-8093 Z\"urich, Switzerland}
\affiliation{Paul Scherrer Institut, CH-5232 Villigen PSI, Switzerland}

\date{\today}

\begin{abstract}
Nuclear magnetic resonance (NMR) measurements of the ${}^{29}$Si spin-lattice 
relaxation time $T_1$ were used to probe the spin-\nicefrac{1}{2} random 
Heisenberg chain compound BaCu$_2$(Si$_{1-x}$Ge$_{x})_2$O$_7$. Remarkable 
differences between the pure ($x=0$) and the fully random ($x=0.5$) case are 
observed, indicating that randomness generates a distribution of local magnetic 
relaxations. 
This distribution, which is reflected in a stretched exponential NMR relaxation, 
exhibits a progressive broadening with decreasing temperature, caused by a 
growing inequivalence of magnetic sites.
Compelling independent evidence for the influence of randomness is also obtained 
from magnetization data and Monte Carlo calculations.
These results suggest the formation of random-singlet states in this class of 
materials, as previously predicted by theory.
\end{abstract}



\maketitle

Random variations of the exchange coupling constant $J$ in ferro-
and antiferromagnetic systems (bond randomness) can have a profound
effect on their magnetic properties. An innovative scheme for studying
random spin-$\nicefrac{1}{2}$ Heisenberg chains (RHCs) was developed 
already in 1979 by Dasgupta and Ma \cite{Dasgupta79}. Their real-space 
renormalization group (RSRG) method was extended by Fisher \cite{Fisher94}, 
and later employed for a much larger variety of problems related to quenched 
disorder in quantum magnets. 
Among these are chains with random ferro- and antiferromagnetic couplings
\cite{Westerberg96}, disordered spin-1 chains and
spin-$\nicefrac{1}{2}$ ladders \cite{Monthus98,Melin02}, Heisenberg
magnets in two and three dimensions \cite{Lin03}, as well as dilute spins 
in doped semiconductors \cite{Bhatt81}. At the core of the RSRG method 
is the so-called decimation procedure. For RHCs it
involves an iterative suppression of the degrees of freedom of the
most strongly coupled spins via the formation of ``frozen''
singlets. Fisher \cite{Fisher94} has shown that this procedure
converges to a \emph{universal} fixed point. The resulting ground
state, characterized by spins coupled at all possible distances and
energy scales, is also known as the \emph{random singlet} (RS)
state. At low temperatures, only excitations of the most weakly
bound singlets are expected to be relevant \cite{Motrunich}. As a result,
regardless of the microscopic details of the disorder, a large class
of bond-disordered systems share universal low-temperature and 
low-energy properties. In recent years, besides providing
predictions for bulk properties, new theoretical work has started to
address also the problem of dynamical observables \cite{Motrunich,Yusuf2005}.

To date only a few good realizations of RHCs have been achieved. 
Some early examples \cite{Tippie81} were later found to have additional 
intricacies beyond the Heisenberg chain model \cite{Sandvik94}.
Usually RHCs are obtained as solid solutions of two prototypical spin-chain 
compounds with different physical properties. A well known example is 
Sr$_3$CuPt$_{1-x}$Ir$_{x}$O$_6$ \cite{Nguyen96}, whose bulk
properties are in agreement with the RSRG predictions. To obtain a 
direct {\it microscopic} insight on the problem, we employed NMR 
experiments to study the local magnetic relaxations in a similar 
system, namely  \BASGE\ \cite{Yamada2000}. 
In this Letter we show that a strongly temperature-dependent distribution of
magnetically inequivalent local relaxations indeed exists in an RHC material. 
In the context of the above-mentioned decimation procedure, we regard this 
phenomenon as an indication of the formation of an RS state.

\begin{figure}[!thbp]
\centering
\includegraphics[width=0.4\textwidth]{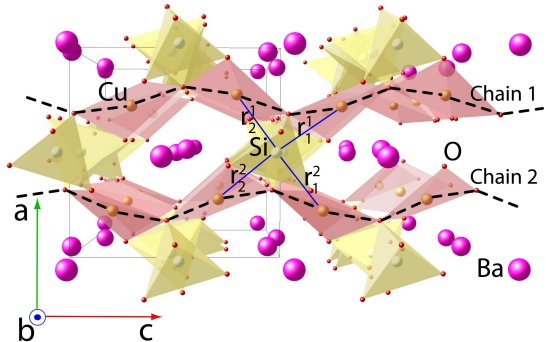} 
\caption{\label{fig:BCSO_struct}(Color online) Crystal structure of \BACSO\ emphasizing 
the copper chains  (dashed lines) and the NN configuration of the silicon atoms.}
\end{figure}

The material BaCu$_2$(Si$_{0.5}$Ge$_{0.5})_2$O$_7$ considered here,
is isostructural with its parent Si- and Ge- compounds. The quasi-1D 
system \BACSO\ (see Fig.~\ref{fig:BCSO_struct}) has an in-chain 
coupling $J = 24.1$ meV \cite{Zheludev01} and orders antiferromagnetically 
at $T_N = 9.2$ K \cite{Tsukada99}. Bond randomness in \BASGE\ is introduced 
via the different bond angles of Cu-O exchange paths, a direct 
consequence of the different Si and Ge covalent radii. A previous study 
\cite{Masuda2004} revealed a logarithmic temperature dependence of the 
spin susceptibility $\chi(T)$ at low temperatures \cite{Dasgupta79}, a 
characteristic RHC feature predicted to be \emph{independent} of the 
initial $J$ distribution. Despite that, and following some original
misinterpretation, inelastic neutron scattering experiments were
found to be quantitatively consistent with Luttinger-liquid
behavior expected for a disorder-free spin chain with a single
effective coupling $J_{\mathrm{eff}} = (J_{\mathrm{Si}} +
J_{\mathrm{Ge}})/2 \approx 37$ meV \cite{Masuda2004,Zheludev07}. 
It was concluded that the RS-related physics manifests itself at 
energies much lower than those accessed by neutron studies. 
Probing these lower-energy/longer-time scales with NMR was the 
main motivation of the present work.

High-quality single crystals of BaCu$_2$(Si$_{1-x}$Ge$_{x})_2$O$_7$ 
with $x=0$ and $x=0.5$ were grown by the floating zone technique. The magnetic 
susceptibility was measured via dc SQUID magnetometry. 
Spin-lattice relaxation times $T_1$ were measured using standard pulse
techniques. Due to the rather long relaxation times (tens of seconds at 
low temperatures), an aperiodic saturation recovery with an echo detection 
was chosen.
With a field $\mu_0 H = 7.024$ T applied along the $a$-axis, the ${}^{29}$Si 
NMR signal was found at 
59.414 MHz.

\begin{figure}[bth]
\centering
\includegraphics[width=0.48\textwidth]{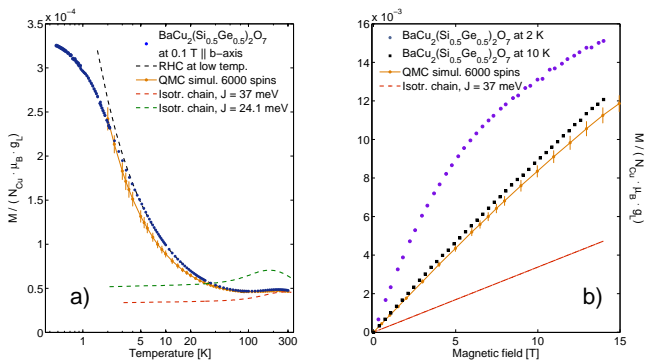} 
\caption{(Color online) Magnetization $M$ per Cu ion of \BASGE\ as a function 
of temperature (a), and field (b), normalized with respect to the saturation 
moment, assuming $g_L = 2$. The fit of $M(T)$ with an RHC model was adopted 
from Ref.~\onlinecite{Fisher94}, while the $M(T)$ and $M(H)$ curves for chains 
with a single isotropic exchange value were calculated as in Ref.~\onlinecite{Johnston2000}. 
QMC simulations of RHCs are also shown (see text). The symbols are explained in the panels.}
\label{fig:squid_fit}
\end{figure}

That it is precisely the low-energy dynamics of \BASGE\
which is most affected by randomness has a clear confirmation 
in the measured bulk magnetic response (Fig.~\ref{fig:squid_fit}). 
Unlike for disorder-free spin chains (light dashed lines), 
the magnetization $M(T)$ of the disordered material (dots) 
tends to diverge at low temperatures.
This divergence is notably different from that observed for $x = 1$, 
where it is related to weak ferromagnetism and thus strongly 
orientation-dependent \cite{Tsukada05}. 
In our case, instead, the $M(T)$ curves only reflect the weak anisotropy of 
the $g$-factor tensor.
We interpret the $x=0.5$, low-$T$ magnetization by the RS theory,
which attributes it to the low-energy states associated with the 
above-mentioned weakly-bound singlets.
It predicts $\chi(T) \sim T^{-1} \ln^{-2}(\varOmega/T)$ 
\cite{Fisher94}, providing a good fit to the data above 4 K with a 
cutoff $\varOmega \simeq 66.3 \pm 0.7$ meV (dark dashed line). 
Since the iterative decimation procedure suppresses the strongly bound 
singlets with an effective coupling larger than $T$, the fraction of 
unpaired spins is $n_T \sim 1/ [\ln(\varOmega/T)]^2$. 
The paramagnetism of this fraction implies $\chi(T) \sim n_T / T$, 
with $\varOmega$ the temperature above which the entire chain 
is effectively in a paramagnetic regime.
The characteristic energy scale of the disorder-induced excitations is, 
however, much lower than $\varOmega$, because already a 5~T magnetic 
field seems to saturate them. This is clearly evidenced by the slope of 
$M(H,2\,\mathrm{K})$, which at higher fields tends to that of a disorder-free chain 
(Fig.~\ref{fig:squid_fit}b).
Quantum Monte Carlo (QMC) simulations of Heisenberg spin-$\nicefrac{1}{2}$ chains, 
using 6000 spin sites with randomly distributed but equally probable $J_{\mathrm{Si}}$ 
and $J_{\mathrm{Ge}}$ couplings, confirm these results. 
$M(T)$ and $M(H)$ averaged over many ($N>40$) random configurations were obtained 
via a directed-loop algorithm \cite{Alet2005} within the ALPS 2.0 package \cite{Bauer2007} 
and are shown in Fig.~\ref{fig:squid_fit}. The $M(H)$ comparison is made at 10 K, where 
the system can be treated as 1D.
Considering the lack of any free parameters, the agreement with data is remarkably good.
Incidentally, a fit of the RS prediction to the low-$T$ QMC results provides 
$\Omega \approx J_{\mathrm{Ge}}$, thus reinforcing the pertinence of the RS theory.

\begin{figure*}[bht]
\centering
\includegraphics[angle=270,width=1\textwidth]{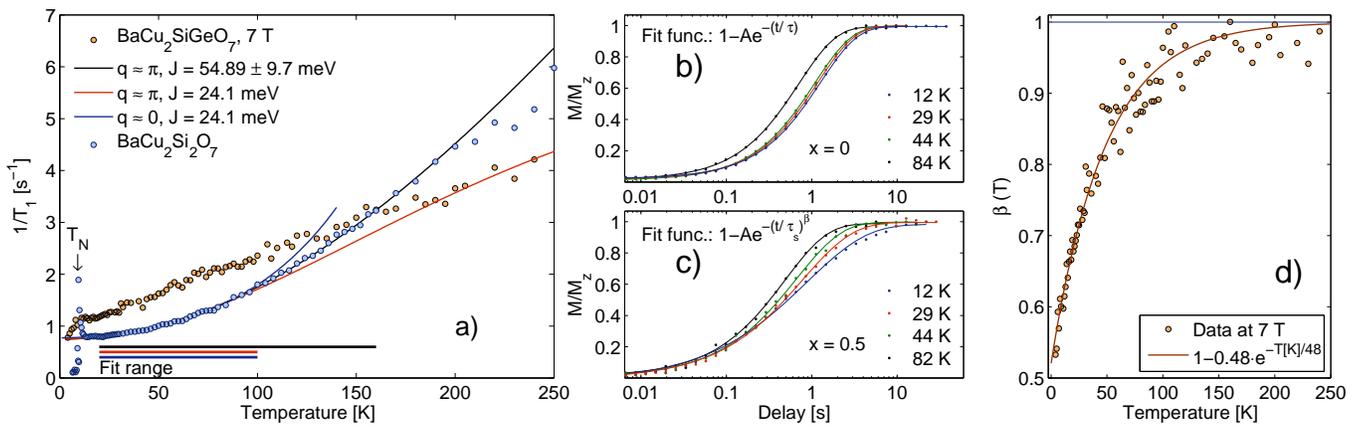} 
\caption{(Color online) ${}^{29}$Si NMR spin-lattice relaxation data of 
Ba\-Cu$_2$\-(Si$_{1-x}$\-Ge$_{x})_2$\-O$_7$ for $x=0$ and $x=0.5$ (no and maximum 
randomness, respectively). a) relaxation rates for both cases, b), c) recovery of 
magnetization, and d) $T$-dependence of $\beta$ for $x=0.5$. The symbols are 
explained in the panels.}
\label{fig:Results}
\end{figure*}

Similarly to the magnetization data, also ${}^{29}$Si NMR relaxation rates 
$T_1^{-1}$ for $x=0$ and $x=0.5$ show remarkable differences 
(see Fig.~\ref{fig:Results}).
The local magnetic field at the silicon nuclei originates from the Cu electronic 
spins $\boldsymbol{S} _j$ and the relevant hyperfine interactions (see Fig.~\ref{fig:BCSO_struct}). 
Provided a matrix description of the latter is adequate and the hyperfine coupling tensor 
at the $j$-site for the $c$-chain is $A^{j,c}$ (see Fig.~\ref{fig:BCSO_struct} for the index notation), 
the $q$-space quantities
$\tilde{A}^{c}_{\alpha, \beta} (\boldsymbol{q})= \sum_{j=1,2} e^{i \boldsymbol{q} \cdot \boldsymbol{r}_j^c} A^{j,c}_{\alpha, \beta}$,
with ${\alpha, \beta} = \{ x,y,z\}$ determine the longitudinal relaxation rate:
\begin{align}
& \frac{1}{T_{1z}} \propto \hbar^2 k_{\mathrm{B}} T  \gamma^2_n
\sum_{\substack{\alpha=x,y,z \\ \boldsymbol{q}}}
\left[ \tilde{A}^2_{x \alpha}(\boldsymbol{q}) + \tilde{A}^2_{y \alpha}(\boldsymbol{q}) \right] \frac{\chi''^c_{\alpha \alpha} (\boldsymbol{q},\omega_n)}{\omega_n} \label{masterIndex}
\end{align}
where $\omega_{n}$ is the resonance frequency.
Detailed calculations reported elsewhere \cite{Casola10} show that the $\boldsymbol{q}$-dependent
prefactor in Eq.~(\ref{masterIndex}), also termed the NMR \emph{form factor},
can be expressed as
$f_{\alpha}^2(\boldsymbol{q}) = M_{\alpha} + P_{\alpha} \cos \left[ \boldsymbol{q} \cdot (\boldsymbol{r}_2-\boldsymbol{r}_1) \right]$,
with $M_{\alpha}$ and $P_{\alpha}$ as constants. 

The $\boldsymbol{q}$-sum in (\ref{masterIndex}) was evaluated numerically 
using the dynamical susceptibility of an isotropic spin-\nicefrac{1}{2}
Heisenberg chain given by theory \cite{Schulz86}. 
Because of the large exchange coupling 
($J = 24.1\,\mathrm{meV} \gg \mu_{\textrm{B}} B \approx 0.4\,\mathrm{meV}$),  
spin chains in \BACSO\ were considered as non-interacting 
above the ordering temperature and effectively in zero field.
Employing the simplest linear description of a Luttinger liquid, i.e., 
without any scattering of quasiparticles and neglecting the influence 
of the form factor, the dominant contribution in (\ref{masterIndex}) is 
at $q = \pi$ and $T_1$ is temperature independent. 
It turns out, however, that including $f_{\alpha}^2(\boldsymbol{q})$ is essential to obtain a 
reasonable fit to the $T_1^{-1}$ data for $x=0$. Considering $f_{\alpha}^2(\boldsymbol{q})$ 
results in an increase of  $T_1^{-1}$ with $T$, as observed and shown in Fig.~\ref{fig:Results}. 
A similar increase, without the influence of  $f_{\alpha}^2(\boldsymbol{q})$, may also be due to
a non vanishing scattering of quasiparticles, implying that the dominant contribution is at $q=0$, 
as shown in Ref.~\onlinecite{Affleck09}.
A convincing demonstration of the different implications of $q=0$ and $q=\pi$ excitations on the 
temperature variations of  $T_1^{-1}$ was given in Ref.~\onlinecite{Thurber01} for Sr$_2$CuO$_3$. 
Additional data of  $T_1^{-1}$  with different crystal orientations (not shown here because of space
restrictions) confirm that $f_{\alpha}^2(\boldsymbol{q})$ cannot be neglected in our case. This 
fact and the verified absence of significant spin-diffusion effects on the field dependence 
of $T_1^{-1}$ indicate that the dominant component is at $q=\pi$. In Fig.~\ref{fig:Results} 
we display the experimental results together with fits considering the two possibilities 
mentioned above. 
Including $f_{\alpha}^2(q)$ and presuming the zero-field limit, only $J$, $M$, and $P$ 
(the latter two independent of $\alpha$), are required as free parameters for the fits.
The best results are obtained with $J \simeq 55$ meV, distinctly larger than expected. 
We recall, however, that the evaluation of $J$ via NMR is not straightforward. It enters 
via a non-zero form factor and its uncertainty depends on the choice of the temperature 
range of the fit. 
Nevertheless, the results for the $x=0$ case show that $^{29}$Si NMR clearly 
probes the Luttinger-liquid physics, hence providing a convenient technique for 
studying the low-energy excitations also for $x>0$.

Referring to the latter case, we notice that due to the loss of translational 
invariance in disordered systems, Eq.~(\ref{masterIndex}) is not valid.
The translational inequivalence of sites introduces a multitude of 
\emph{local} nuclear relaxation times. For a nuclear spin $I=\nicefrac{1}{2}$, 
the individual relaxations are simply exponential, but collectively they give 
rise to a system-dependent and generally non-exponential relaxation.
If $\tau$ is one of the local relaxation times, a possible interpretation 
for the observed stretched exponential recovery is provided by \cite{Montroll,Lindsey}:  
\begin{equation}
\int_{0}^{\infty} \!\! \rho(\tau,T)\; e^{-\frac{t}{\tau(T)}} \; \diff \tau =
e^{-\left[ \frac{t}{\tau_s (T)} \right]^{\beta (T)}}.
\label{stretched}
\end{equation}
This equation implies that if the same type of relaxation occurs with a probability 
$\rho (\tau,T)$ and with a characteristic 
temperature-dependent $\tau(T)$ at different nuclear sites, then the global 
relaxation function is a stretched exponential, characterized by 
the parameters $\beta(T)$ and $\tau_s(T)$. \\
Relaxations described by a stretched exponential are quite common in disordered magnets.
The temperature dependence of the exponent $\beta$ has been reported for a number of 
different systems \cite{Phillips}, such as spin glasses, quasicrystals, polymers, etc.
In our case, while for $x=0$ a simple exponential is adequate for describing 
the magnetization recovery, stretched exponentials are needed for $x=0.5$. 
The corresponding data and fits are shown in Fig.~\ref{fig:Results}b and 
\ref{fig:Results}c, respectively. 
From the latter, $\tau_s(T) \equiv T_1$ and $\beta(T)$ are obtained and shown in 
Figs.~\ref{fig:Results}a and \ref{fig:Results}d.

\begin{figure}[!hbtp]
\centering
\includegraphics[width=0.48\textwidth]{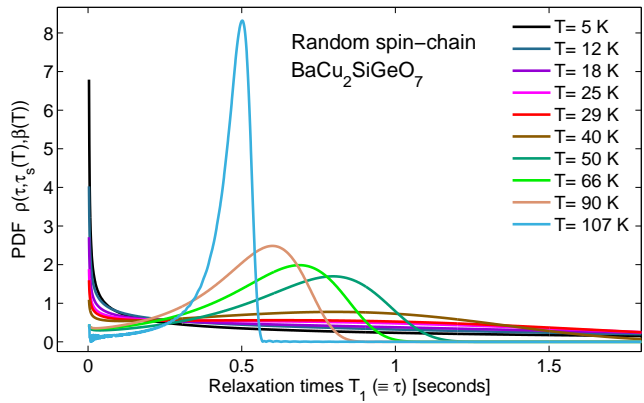} 
\caption{(Color online) Probability distribution of NMR spin-lattice re\-la\-xa\-tion 
times in 
\BASGE\ 
at different temperatures, as derived from 
fits to the experimental data partly shown in Fig.~\ref{fig:Results}c.}  
\label{fig:Distribution}
\end{figure}

Extracting the weights $\rho (\tau,T)$ via an inversion of Eq.~(\ref{stretched}) 
is a mathematically non-trivial task \cite{Lindsey}. 
To account for our case, where $\beta \geq 0.5$, we start from Eq.~\eqref{stretched}, 
but now using the variable $s = \tau_s/\tau$ \cite{Johnston}:
\begin{equation}
e^{-(t/\tau_s)^{\beta}} = \int_{0}^{\infty}\!\! P(s,\beta) \;e^{-s t/\tau_s} \, \diff s,
\label{startJohn}
\end{equation}
where $P(s,\beta)$, as $\rho(\tau)$ before, represents a particular probability 
distribution function (PDF) of relaxation rates for a given value of $\beta$.
The inverse of the Laplace transform of \eqref{startJohn} 
\cite{Johnston} 
%
%
provides the required relation between $P$ and $\rho$, the probability 
distributions of relaxation \textit{rates} and \textit {times}, respectively:
\begin{equation}
\rho \left( \frac{\tau}{\tau_s}, \beta \right) = P \left( \frac{\tau_s}{\tau}, \beta \right) \frac{\tau_s}{\tau^2}.
\label{finalSub}
\end{equation}
Although $P(s,\beta)$ is obtained only numerically, Eq.~(\ref{finalSub}) allows us to 
extract the distribution of the local relaxation times in a random Heisenberg chain 
starting from the experimental NMR data.
Results of this procedure for a series of different temperatures are shown in Fig.~\ref{fig:Distribution}.
\noindent
%
At high temperatures, where $\beta \rightarrow 1$, the distribution of relaxation times 
tends to a delta function peaked at $\tau = \tau_s(T)$, while at low temperatures the 
distribution diverges weakly as $\tau \rightarrow 0$ and hence remains 
integrable \cite{Lindsey}. 
%
In all cases the distribution is normalized and does not depend 
on temperature, thus reflecting the conserved total number of magnetic sites.
In a disordered system the sites are magnetically inequivalent, 
giving rise to a broadening of the relaxations' distribution at low temperatures. 
For $T \rightarrow 0$, the distribution $\rho(\tau,\beta)$ peaks at $\tau \rightarrow 0$,
but its mean value $\tau_s$ increases due to the longer tail of the distribution.

We argue that the measured NMR parameters can be used to obtain information 
on the low-energy \textit{dynamics} of RHC systems, inaccessible by standard neutron-scattering 
techniques \cite{Masuda2004}. In case of ran\-dom\-ness, the lack of translational 
invariance prompts for models based on real- instead of reciprocal-space configurations. 
In the RS picture, upon lowering the temperature there is an increase in the number of 
singlets with different couplings being formed, while $n_T$ goes to zero.
This results in a continuous distribution of magnetically inequivalent sites and 
in longer $\tau_s(T)$ at lower temperatures, as reflected in our data.
In this low-$T$ scenario the low-energy dynamics is determined by the 
increasing importance of the ``frozen'' singlets, at the expense of the yet 
uncoupled spins.
We shall address this point in a forthcoming publication \cite{Casola10}.

In conclusion, we have carried out a detailed ${}^{29}$Si NMR study of the spin-lattice 
relaxation rates in the BaCu$_2$(Si$_{1-x}$Ge$_x$)$_2$O$_7$ system for $x=0$ and $x=0.5$, 
corresponding to no and to maximum disorder, respectively.
NMR turns out to be an appropriate technique for the study of low-energy dynamics in 
gapless antiferromagnets with bond randomness.
Guided by the experimental observation that the relaxation of magnetization is 
characterized by a stretched exponential, we show that it is possible to extract 
the distribution of relaxation functions in a random Heisenberg chain from 
experiment. 
With the additional evidence from magnetization data and QMC calculations, 
we interpret this result to reflect the formation of random-singlet states in a random 
Heisenberg chain, as predicted by theory.

The BaCu$_2$SiGeO$_7$ crystals used in this work were prepared by Dr.\ T.\ Masuda 
in the group of prof.\ K.\ Uchinokura at the University of Tokyo. 
The authors thank T.\ Giamarchi and P.\ Bouillot for useful suggestions concerning 
the NMR form factor. This work was financially supported in part by the Schweizerische 
Nationalfonds zur F\"{o}rderung der Wissenschaftlichen Forschung (SNF) and the 
NCCR research pool MaNEP of SNF.


\begin{thebibliography}{99}
%
\bibitem{Dasgupta79}
C. Dasgupta and S.-K. Ma, Phys. Rev. B \textbf{22}, 1305 (1980).
%
\bibitem{Fisher94}
D. S. Fisher, Phys. Rev. B \textbf{50}, 3799 (1994).
%
\bibitem{Westerberg96}
E. Westerberg, A. Furusaki, M. Sigrist, and P. A. Lee, Phys. Rev. B \textbf{55}, 12578 (1997).
%
\bibitem{Monthus98}
R. A. Hyman and K. Yang, Phys. Rev. Lett. \textbf{78}, 1783 (1997); C. Monthus, O. Golinelli, and Th. Jolicoeur, Phys. Rev. B \textbf{58}, 805 (1998).
%
\bibitem{Melin02}
R. M\'elin, Y.-C. Lin, P. Lajk\'o, H. Rieger, and F. Igl\'oi, Phys. Rev. B, \textbf{65}, 104415 (2002).
%
\bibitem{Lin03}
Y.-C. Lin, R. M\'elin, H. Rieger, F. Igl\'oi, Phys. Rev. B \textbf{68}, 024424 (2003).
%
\bibitem{Bhatt81}
R. N. Bhatt and P. A. Lee, Phys. Rev. Lett. \textbf{48}, 344 (1982).
%
\bibitem{Motrunich}
O. Motrunich, K. Damle, and D. A. Huse, Phys. Rev. B \textbf{63}, 134424 (2001).
%
\bibitem{Yusuf2005}
E. Yusuf and K. Yang, Phys. Rev. B \textbf{72}, 020403(R) (2005).
%
\bibitem{Tippie81}
L. C. Tippie and W. G. Clark, Phys. Rev. B \textbf{23}, 5846 (1981); \textbf{23}, 5854 (1981).
%
\bibitem{Sandvik94}
A. W. Sandvik, D. J. Scalapino, and P. Henelius, 
Phys. Rev. B \textbf{50}, 10474 (1994).
%
\bibitem{Nguyen96}
T. N. Nguyen, P. A. Lee, H.-C. zur Loye, Science \textbf{271}, 489 (1996).
%
\bibitem{Yamada2000}
Y. Yamada, Z. Hiroi, and M. Takano, J. Solid State Chem. \textbf{156}, 101 (2000).
%
\bibitem{Zheludev01}
A. Zheludev et al. 
Phys. Rev. B \textbf{65}, 014402 (2001).
%
\bibitem{Tsukada99}
I. Tsukada et al. 
Phys. Rev. B \textbf{60}, 6601 (1999).
%
\bibitem{Masuda2004}
T. Masuda et al. 
Phys. Rev. Lett. \textbf{93}, 077206 (2004); \textbf{96}, 169908(E) (2006).
%
%
\bibitem{Zheludev07}
A. Zheludev et al. 
Phys. Rev. B \textbf{75}, 054409 (2007).
%
\bibitem{Johnston2000}
D. C. Johnston et al.,
Phys. Rev. B \textbf{61}, 9558 (2000).
%
\bibitem{Tsukada05}
I. Tsukada, J. Takeya, T. Masuda, and K. Uchinokura, Phys. Rev. B \textbf{62} (2000) R6061.
%
\bibitem{Alet2005}
F. Alet, S. Wessel, and M. Troyer, Phys. Rev. E \textbf{71}, 036706 (2005).
%
\bibitem{Bauer2007}
B. Bauer et al. (ALPS collaboration) arXiv:1101.2646 [cond-mat].
%
\bibitem{Casola10}
F.\ Casola \emph{et al.}, (to be published).
%
\bibitem{Schulz86}
H. J. Schulz, Phys. Rev. B \textbf{34}, 6372 (1986). 
%
%
\bibitem{Affleck09}
J. Sirker, R. G. Pereira, and I. Affleck, Phys. Rev. Lett. \textbf{103}, 216602 (2009).
%
\bibitem{Thurber01}
K. R. Thurber, A. W. Hunt, T. Imai, and F. C. Chou, Phys. Rev. Lett. \textbf{87}, 247202 (2001).
%
\bibitem{Montroll}
E. W. Montroll and J. T. Bendler, J. of Stat. Phys. \textbf{34}, 129 (1983).
%
\bibitem{Lindsey}
C. P. Lindsey and G. D. Patterson, J. Chem. Phys. \textbf{73}, 3348 (1980).

\bibitem{Phillips}
J. C. Phillips, Rep. Prog. Phys. \textbf{59}, 1133 (1996).
%
\bibitem{Johnston}
D. C. Johnston, Phys. Rev. B \textbf{74}, 184430 (2006).
%

\end{thebibliography}
\end{document}